\documentclass[useAMS,usenatbib]{mnras}

\usepackage{tabularx}
\usepackage{multirow}
\usepackage{url}
\usepackage{amsmath}	
\usepackage{amssymb}	
\usepackage{aas_macros}
\usepackage{graphicx}
\usepackage{times,txfonts}
\usepackage{bm}

\usepackage[T1]{fontenc}
\usepackage{ae,aecompl}
\newcommand\lb{\langle}
\newcommand\rb{\rangle}

\title[Rayleigh unstable MHD shear turbulence]{Shearing box simulations in the Rayleigh unstable regime}
\author[F. Nauman and E. G. Blackman]{Farrukh Nauman$^{1,2}$\thanks{E-mail:
nauman@nbi.ku.edu} and Eric G. Blackman $^{1,3}$\thanks{E-mail: blackman@pas.rochester.edu}\\
$^{1}$Department of Physics and Astronomy, University of Rochester, Rochester, NY 14627, USA.\\
$^{2}$Niels Bohr International Academy, The Niels Bohr Institute, Blegdamsvej 17, DK-2100, Copenhagen \O, Denmark.\\
$^{3}$School of Natural Sciences, Institute for Advanced Study, Princeton, NJ 08540, USA.}

\pubyear{2015}

\begin{document}

\date{\today}

\pagerange{\pageref{firstpage}--\pageref{lastpage}}

\maketitle

\label{firstpage}

\begin{abstract}
We study the stability properties of Rayleigh unstable flows both in the purely hydrodynamic and magnetohydrodynamic (MHD) regimes for two different values of the shear $q=2.1, 4.2$ ($q = - d\ln\Omega / d\ln r$) and compare it with the Keplerian case $q=1.5$. We find that the $q>2$ regime is unstable both in the hydrodynamic and in the MHD limit (with an initially weak magnetic field). In this regime, the velocity fluctuations dominate the magnetic fluctuations. In contrast, in the $q<2$ (magnetorotational instability (MRI)) regime the magnetic fluctuations dominate. This highlights two different paths to MHD turbulence implied by the two regimes, suggesting that in the $q>2$ regime the instability produces primarily velocity fluctuations that cause magnetic fluctuations, with the causality reversed for the $q<2$ MRI unstable regime. We also find that the magnetic field correlation is increasingly localized as the shear is increased in the Rayleigh unstable regime. In calculating the time evolution of spatial averages of different terms in the MHD equations, we find that the $q>2$ regime is dominated by terms  which are nonlinear in the fluctuations, whereas for $q<2$, the linear terms play a more significant role.
\end{abstract}


\section{Introduction}\label{sec:intro}
Differentially rotating flows are ubiquitous in astrophysics and studying their stability has been a long-standing enterprise. Using the local shearing box approximation (\cite{1965MNRAS.130...97G}, \cite{1995ApJ...440..742H} with Keplerian shear ($q=1.5$), numerical simulations have shown that the Magnetorotational Instability (MRI) leads to turbulent growth of stresses in the presence of a weak magnetic field (for example, \cite{1959velikhov}, \cite{1960PNAS...46..253C}, \cite{1991ApJ...376..214B}). The Rayleigh criterion, based on a linear modal analysis of axisymmetric perturbations, suggests that Keplerian flow is stable in hydrodynamics. This, however, does not rule out the possibility of subcritical transition to turbulence (\cite{1996ApJ...467...76B}, \cite{2005A&A...444...25L}). 

The (Rayleigh stable) Keplerian flow has understandably received the most attention because of its direct application in accretion discs, but here we focus on the stability properties of hydrodynamic and magnetohydrodynamic (MHD) flow in the Rayleigh unstable regime $q>2$. A study of the Rayleigh unstable regime is of interest because a comprehensive understanding of shear driven MHD turbulence requires knowing the differences in the $q<2$ and $q>2$ regimes. Additionally, certain astrophysical flows are actually thought to be Rayleigh unstable. These include counter rotating accretion discs (e.g., \cite{2015MNRAS.446..613D}), counter rotating galaxies (e.g., \cite{2014ASPC..486...51C}) and the plunging region close to a black hole (e.g., \cite{1978A&A....63..221A}, \cite{1996MNRAS.281L..21A}, \cite{2004ApJ...614..309G}, \cite{2012MNRAS.423L..50B}, \cite{2013MNRAS.428.2255P}).

While the standard shearing box in the Rayleigh unstable regime poses challenges that we discuss further in section \ref{sec:rayl}, certain properties of the shear instabilities in both the hydrodynamic and magnetohydrodynamic (MHD) case can be studied numerically with an appropriate configuration and code. Toward this end, we have conducted numerical simulations for three different values of $q$ ($1.5, 2.1, 4.2$) both in pure hydrodynamics and MHD. We first used the publicly available finite volume code \textsc{athena} \footnote{\url{https://trac.princeton.edu/Athena/}} ((\cite{2005JCoPh.205..509G}, \cite{2008ApJS..178..137S}, \cite{2010ApJS..189..142S}) and found that even though we started out with zero initial  momenta, truncation errors introduced perturbations that led to the exponential growth of the mean momentum  and the eventual crash of the simulation (the time step is inversely proportional to maximum velocity). We then chose the pseudospectral code \textsc{snoopy} \footnote{\url{http://ipag.osug.fr/~lesurg/snoopy.html}} (\cite{2005A&A...444...25L}, \cite{2011A&A...528A..17L}) to simulate $q>2$, which conserves the $k=0$ mode. 

In section \ref{sec:stab} we review the linear stability theory of hydrodynamic and magnetohydrodynamic shear flows and  discuss it in the context of shearing box approximation. In section \ref{sec:results}, we describe the numerical setup and simulation results. We conclude in section \ref{sec:conclusions}.

\section{Stability of shear flows}\label{sec:stab}

\subsection{Linear analysis}\label{sec:lin}
Following the discussion in \cite{2015MNRAS.448.3697S}, the dispersion relation for local axisymmetric perturbations of the form $e^{i(\omega t - k_r r - k_z z)}$  (see also \cite{2012MNRAS.423L..50B} for the special case of $k_r=0$) with the initial magnetic field $B_0$ pointing in the $z$ direction is (\cite{1959velikhov}, \cite{1960PNAS...46..253C}, \cite{1991ApJ...376..214B}, \cite{1998bhad.conf.....K}, \cite{2015MNRAS.448.3697S}):
\begin{equation}
\omega^4 - \omega^2 \left(2 k_z ^2 v_A^2 + \left(\frac{k_z}{k}\right)^2 \kappa^2 \right) + k_z^2 v_A^2 \left(k_z^2 v_A^2 + \left(\frac{k_z}{k}\right)^2 \kappa^2 - 4 \Omega ^2 \right) = 0,
\label{eq:dispersion}
\end{equation}
where $k^2 = k_r^2 + k_z^2$, $\kappa^2 = 4\Omega^2 + r d \Omega^2/dr = 2 \Omega^2 (2 - q)$, $v_A^2 = B_0^2/(4\pi \rho_0)$ and $\rho_0 = \text{initial density}$. The solution is\begin{equation}
\omega^2 = \left(\frac{k_z}{k}\right)^2  \left(  k^2 v_A^2 + \frac{\kappa^2}{2} \pm \sqrt{ \frac{\kappa^2}{4} + 4 \Omega^2 k^2 v_A^2 }  \right).
\label{eq:dispsol}
\end{equation}
For the classical Rayleigh criterion in hydrodynamics, $v_A = 0$ and the above relation gives $\omega^2 = (k_z/k)^2 \kappa^2$. This implies that purely hydrodynamic perturbations are stable as long as $\kappa^2>0$, or equivalently $q<2$. However, the addition of magnetic fields makes the $q<2$ regime unstable and instead $\omega_{\text{MRI}}^2 \sim (k^2 v_A^2)/(\kappa^2 d\Omega^2/d \ln r) = - q/(2-q) k^2 v_A^2 $ in the limit $k^2 v_A^2 << 1$ \citep{2012MNRAS.423L..50B}.

We focus our attention to the $q>2$ or $\kappa^2<0$ regime in this paper. It is convenient to define the two different branches of Eq. \ref{eq:dispsol} in the limit of $k^2 v_A^2 << 1$ as:
\begin{align}
\omega_R^2 &= \left(\frac{k_z}{k}\right)^2  \left( \kappa^2 + k^2 v_A^2 \left( 1 + \frac{4\Omega^2}{\kappa^2}  \right) \right) \label{eq:rayl} \\
\omega_{VC}^2 &= k_z^2 v_A^2 \left( 1 - \frac{4\Omega^2}{\kappa^2}  \right) \label{eq:vc}
\end{align}
where $\omega_R = \text{Rayleigh mode}$ and $\omega_{VC} = \text{Velikhov-Chandrasekhar mode}$. As explained by \cite{2015MNRAS.448.3697S}, these modes are so named because we recover the classical Rayleigh instability criterion from the Rayleigh mode in the absence of magnetic field ($v_A=0$), and the VC mode vanishes in this limit. In the regime $\kappa^2<0$, it follows from  above that the VC mode is stable for all wavenumbers and only the Rayleigh mode is unstable. This distinction between the Rayleigh and VC mode was not made in \cite{2012MNRAS.423L..50B}.

\subsection{Shearing box in the Rayleigh unstable regime}\label{sec:rayl}
The shearing box approximation in the ideal compressible MHD limit is discussed in \cite{2015MNRAS.446.2102N}. Here we revisit that discussion in the context of non-ideal incompressible MHD equations since \textsc{snoopy} solves this set of equations. The shearing box equations in the frame co-moving with the background shear velocity $\bm{v}_{sh} = - q \Omega x \bm{e}_y$ are:
\begin{gather}
\frac{\partial \bm{v}}{\partial t} + v_{sh} \frac{\partial \bm{v}}{\partial y} + \nabla \cdot (\bm{v} \bm{v} + \bm{T}) = 2 \Omega v_y \bm{e}_x + (2 - q) \Omega v_x \bm{e}_y + \nu \nabla^2 \bm{v}, \label{eq:NS} \\
\frac{\partial \bm{b}}{\partial t} = \nabla \times (\bm{v} \times \bm{b}) + \eta \nabla^2 \bm{b}, \label{eq:induc}\\
\nabla \cdot \bm{v} = 0, \\
\nabla \cdot \bm{b} = 0, 
\end{gather}
where $\bm{v}$ and $\bm{b}$ are the velocity and magnetic field respectively. Here $\bm{T}$ is a stress tensor given by
\begin{equation}
\bm{T} = (p + b^2/2) \bm{I} - \bm{b}\bm{b},
\end{equation}
where $\bm{I}$ is the identity matrix and $p$ is thermal pressure. 

Upon volume averaging the Navier-Stokes equation (eq. ~\ref{eq:NS}), we obtain two coupled equations for the volume averaged velocities $\langle v_x\rangle$ and $\langle v_y\rangle $:
\begin{align}
\frac{\partial \langle v_x \rangle}{\partial t} &= 2 \Omega \langle v_y \rangle, \label{epi1} \\
\frac{\partial \langle v_y \rangle}{\partial t} &= \Omega \langle v_x \rangle (q-2). \label{epi2}
\end{align}
which yields the solution that both averaged velocities are proportional to $\exp(\pm i \kappa t) $ for $q<2$, or $\sim \exp(\pm \kappa t)$ for $q>2$ where $\kappa^2 = 2\Omega^2 (2-q)$. 
The above analysis shows that the `$x$' and `$y$' mean velocities will grow exponentially, if perturbed, in the Rayleigh unstable regime $q>2$. This growth is a physical effect for finite perturbations. However  if we set initial mean velocities to be zero the physical
velocities should remain such, but in simulations they can grow because of truncation errors. We verified this with the finite volume code \textsc{athena}.  The truncation errors seeded the mean velocities and they grew exponentially bringing the simulation to a halt in just a few shear times ($1/(q \Omega)$). 

We therefore chose to use the  publicly available incompressible pseudospectral code \textsc{snoopy}, which has the important property that the box averaged mean velocities do not grow throughout the duration of the simulation. This is because the nonlinear terms in the code are of the form $(i\bm{k} \cdot \bm{v}) \bm{v}$, and do not contribute when $k=0$. Linear terms can only contribute to $k=0$ mode evolution if the initial value for the fields at $k=0$ is not set to zero,
but we started all of our simulations without perturbations in this mode.

\section{Numerical results}\label{sec:results}
\subsection{Setup}\label{sec:setup}
Using \textsc{snoopy}, we solve  the incompressible hydrodynamic and MHD equations in the shearing box approximation. {We solve the equations where the background shear has been subtracted out. \textsc{snoopy} utilizes the the 2/3 antialiasing rule \citep{canuto2006}. Shear periodic boundaries are remapped every $t_{\text{remap}} = L_y/(q\Omega L_x)$ \citep{2004A&A...427..855U}. We define the Reynolds and magnetic Reynolds numbers $Re = L_z^2 q \Omega/\nu$, $Rm = L_z^2 q \Omega/\eta$, respectively, where $L_z=\Omega=1$ in code units. We fixed $Re = Rm = 1600$ for most of our runs.

We use large scale noise as initial perturbations (with zero mean) and set the net initial vertical field $B_0 = 0.025$ in code units, which corresponds to an initial plasma beta $\beta = L_z^2 \Omega^2/ (B_0^2/2) = 3200$. The magnetic field is calculated in Alfven speed units. For all of our runs, we use the domain size $L_x = L_y = L_z = 1$ with a resolution of $64^3$. Table 1 provides a summary of our runs.
\begin{table*}
\centering
\begin{tabular}{| c | c | c | c | c | c | c |}
\hline \hline
  Run & Shear & $\overline{\langle v_xv_y \rangle}$ & -$\overline{\langle b_xb_y \rangle}$ & $\alpha_{\text{kin,y}}\equiv \overline{\langle v_xv_y\rangle/\langle v_y^2\rangle}$ & $\alpha_{\text{mag,y}}\equiv \overline{-\langle b_xb_y \rangle/\langle b_y^2 \rangle}$ \\ \hline \hline
  mhd15 & 1.5 & $0.4837 \pm 0.3812$ & $3.3940 \pm 3.8068$ & $0.4027 \pm 0.1751$ & $0.1877 \pm 0.0257$\\
  mhd21 & 2.1 & $0.5896 \pm 0.5397$ & $0.4824 \pm 0.3851$ & $0.8701 \pm 0.4520$ & $0.2334 \pm 0.0889$\\
  mhd42 & 4.2 & $2.3140 \pm 3.3113$ & $0.0735 \pm 0.0871$ & $0.5987 \pm 0.1660$ & $0.1712 \pm 0.1164$\\ \hline
  hyd21 & 2.1 & $0.0519 \pm 0.0624$ & & $0.7331 \pm 0.2993$ & \\
  hyd42 & 4.2 & $0.9823 \pm 0.9622$ & & $0.5210 \pm 0.1351$ & \\ \hline \hline
\end{tabular}
\caption{The first three runs are MRI runs whereas the last two are the purely hydrodynamic runs. We do not list the Keplerian hydrodynamic run here as it did not become turbulent. All the quantities are time averaged from $1000 (1/\Omega)$ to $2000 (1/\Omega)$ (time averaging is defined by an overline) for all of the runs and volume averaged (represented by angled brackets) over the whole box. The stresses $\langle v_xv_y \rangle$ and $\langle b_xb_y \rangle$ are normalized by $L_z^2 \Omega^2$, which equals unity according to our definitions. The fifth column represents the ratio of the Reynolds stress to the square of the azimuthal velocity $\alpha_{\text{kin,y}}\equiv \overline{\langle v_xv_y\rangle/\langle v_y^2\rangle}$, while the last column shows this ratio corresponding to the magnetic field $\alpha_{\text{mag,y}}\equiv \overline{-\langle b_xb_y \rangle/\langle b_y^2 \rangle}$. It appears that $\alpha_{\text{kin,y}}$ is a sensitive function of the shear parameter while $\alpha_{\text{mag,y}}$ is roughly constant.} 
\label{tab:1}
\end{table*}

\subsection{Hydrodynamic shear flow stability}\label{sec:hydro}
\begin{figure}

  \centering
    \includegraphics[width=0.45\textwidth]{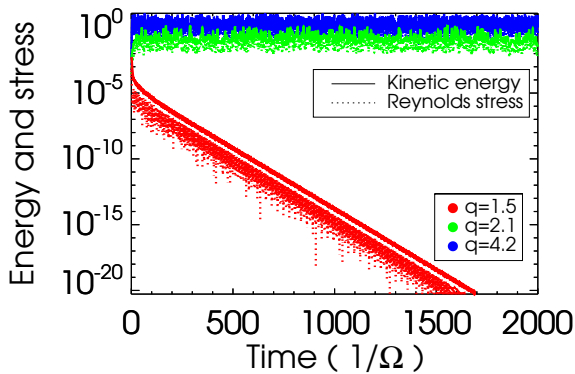}

  \caption{Time history plot of kinetic energy (solid) and Reynolds stress (dotted) for hyd15 ($q=1.5$, red), hyd21 ($q=2.1$, green) and hyd42 ($q=4.2$, blue). The y-axis is in log scale and the x-axis is time in units of $1/\Omega$.}
\label{fig:hydke}
\end{figure}

As discussed in the previous section, the $q<2$ regime is stable in hydrodynamics (see also \cite{FLM:397764}, \cite{1997JFM...347..289B}, \cite{2005JFM...542..305B} for earlier work). We checked this by simulating the Keplerian $q=1.5$ regime as well as two different values of shear in the Rayleigh unstable regime $q=2.1,4.2$. We plot the time history of the kinetic energy and the Reynolds stresses in Fig. ~\ref{fig:hydke}. As predicted by the standard modal analysis, the Keplerian flow is stable and its fluctuations exponentially decay to zero whereas the two Rayleigh unstable runs reach a saturated turbulent state in just a few shear times.

\subsection{MHD shear flow stability}\label{sec:mhd}
For MHD the regime  $0<q<2$ is unstable to the MRI. In \cite{2015MNRAS.446.2102N}, we focused on the dependence on $q$ for  $q<2$  and found that the results were consistent with the linear calculations of \cite{2006MNRAS.372..183P} and the empirical results of \cite{1996MNRAS.281L..21A}. In contrast, the $q>2$ case is stable to the MRI so a comparison of saturated states of the two regimes is instructive.

\begin{figure}

  \centering
    \includegraphics[width=0.45\textwidth]{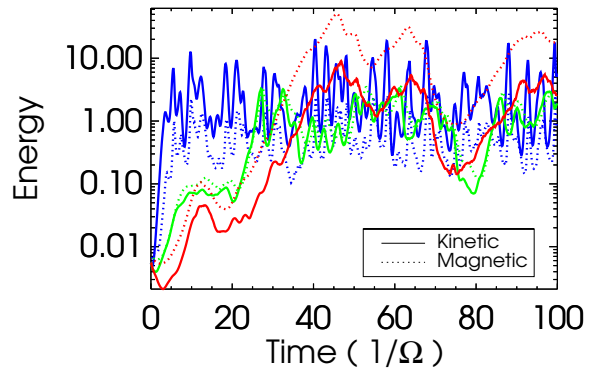}

  \caption{Time history plot of kinetic (solid) and magnetic energies (dotted) for mhd15 ($q=1.5$, red), mhd21 ($q=2.1$, green) and mhd42 ($q=4.2$, blue). }
\label{fig:meke}
\end{figure}

\begin{figure}

  \centering
    \includegraphics[width=0.45\textwidth]{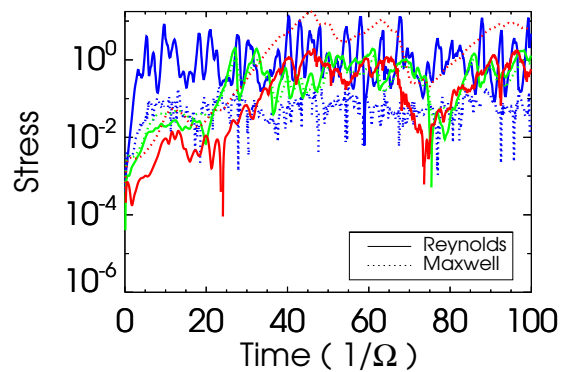}

  \caption{Same as Fig. \ref{fig:meke} but for Reynolds (solid) and Maxwell stresses (dotted).}
\label{fig:mxyrxy}
\end{figure}

One common feature visible from Figs. \ref{fig:hydke}, \ref{fig:meke} and \ref{fig:mxyrxy} is that the case of largest shear (blue line, $q=4.2$) has the largest growth rate in both magnetic and kinetic energies. The trend of increased growth rate with shear is also a property of the $q<2$ ($\kappa>0)$ MRI regime  \citep{2015MNRAS.446.2102N}. However, the important difference to note both in Fig. \ref{fig:meke} and \ref{fig:mxyrxy} is that the growth rate of the kinetic energy (Reynolds stress) is greater than that of magnetic energy (Maxwell stress) in the $q>2$ regime.

To further explore the difference between kinetic and magnetic energy in the $q>2$ regime, we increased $Re$ and $Rm$ to $6400$ and $12800$ (at $Pr_M = Rm/Re = 1$) for $q=4.2$ and observed that the ratio of kinetic energy to magnetic energy in the saturated state decreased to nearly $2.7$ for $Re=Rm=12800$ compared to $\sim 5.0$ for the $Re=Rm=1600$ and $6400$ cases. An extensive study of $Re$, $Rm$ dependence is beyond the scope of the current paper. For Keplerian flow, the turbulent stresses also depend on dissipation coefficients (see for example, \cite{2015A&A...575A..14R}). 

As reviewed in section \ref{sec:rayl} above linear theory suggests that we can break the dispersion relation into two different types of modes \citep{2015MNRAS.448.3697S}: Rayleigh and Velikhov-Chandrasekhar (VC). For $q>2$, the VC mode is stable at all wave numbers. Our results show that for $q<2$, the magnetic energy leads the kinetic energy while for $q>2$ the kinetic energy leads the magnetic energy. This result is reminiscent of isotropically forced box simulations of MHD turbulence in the following sense. In such simulations, the turbulent driver is imposed by hand as a forcing function. Normally the forcing is in the the Navier-Stokes equation, but it can also be imposed in the induction equation. When the forcing is imposed in the Navier-Stokes equation the saturated state reveals that the kinetic energy dominates the magnetic energy at the forcing scale and below. In contrast, when the forcing is in the induction equation, the magnetic energy dominates the kinetic energy at these scales \citep{Park2012}.

These circumstances  reflect the fact that the transfer of energy from the quantity that is driven ($v$ or $b$) is not 100\% efficient to the response quantity ($b$ and $v$, respectively). Interpreted in this way, the results from our simulations suggest that the for the $q>2$ regime, the Rayleigh mode acts more like an  an effective ``driving" in the Navier Stokes equation, whereas for the $q<2$ regime, the VC mode perhaps leads to a kind of ``effective" forcing in the induction equation. This physical distinction may be useful in the path toward constructing analytic theoretical approaches and is consistent with toy models in the MRI context that invoke forcing in the induction equation (e.g. \cite{2015PhRvL.114h5002S}). More work is needed to assess this rigorously.

Finally, we note that boxes that are sufficiently large in the direction normal to the shear ($L_y,L_z \gg L_x$) can lead to qualitatively different regime of `spatiotemporal chaos' (\cite{pomeau},\cite{philip}). For $q<2$ MHD shearing box simulations with $L_z\gg L_x$, \cite{shi2016} showed that coherent structures appear in the magnetic field while more recently \cite{nauman2016} have shown that both velocity and magnetic fields develop coherent structures. The boxes used in the present study have $L_x=L_y=L_z=1$, so the extent to which a similar role of large boxes might also apply to the Rayleigh unstable regime should be investigated in future work.

\subsection{Correlation in space (x-y plane)}\label{sec:corrxy}

\begin{figure*}

  \centering
    \includegraphics[width=1.\textwidth]{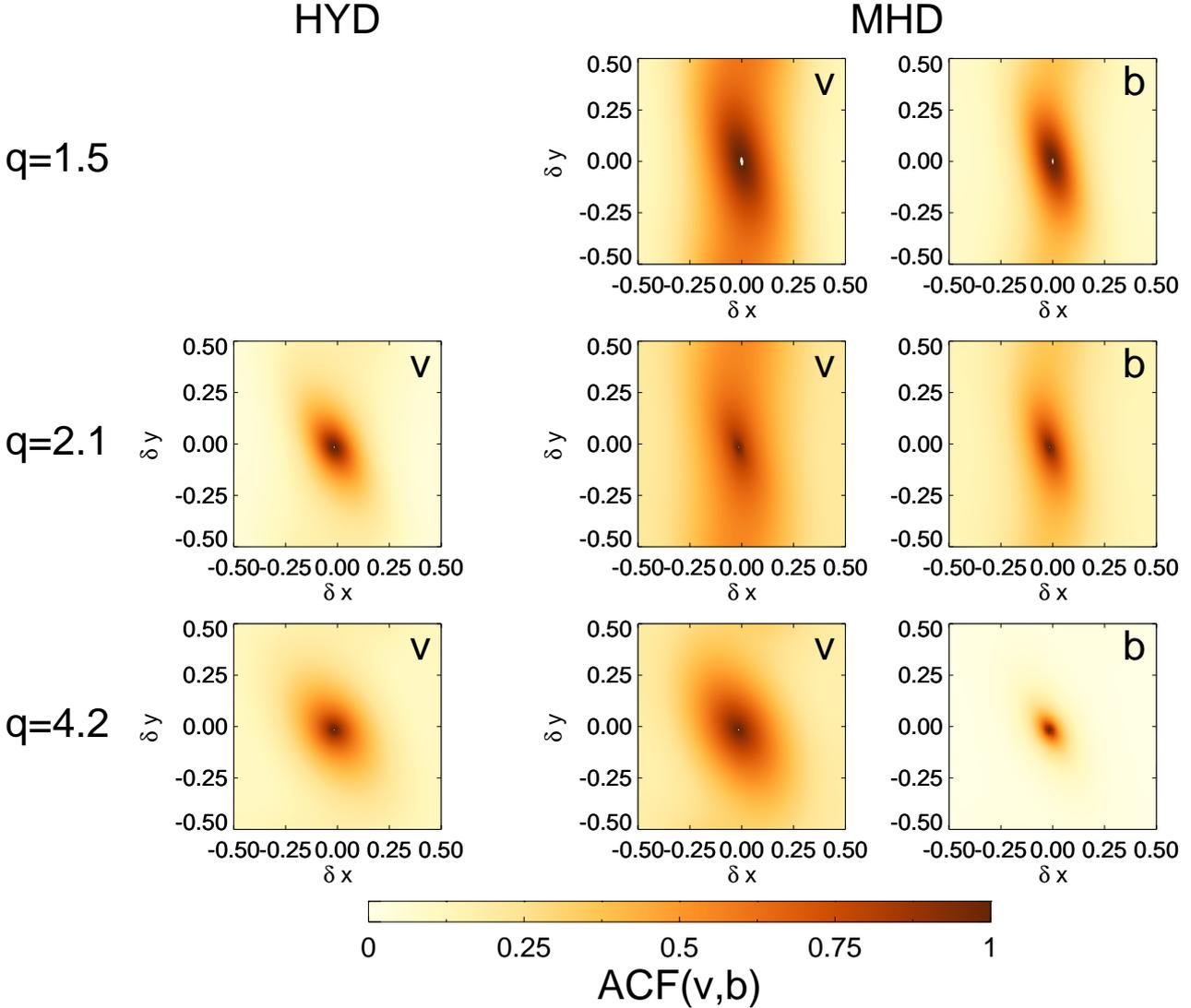}

  \caption{Contour plots of the autocorrelation of velocity and magnetic fields for different runs.}
\label{fig:contour}
\end{figure*}

Studying the physical effect of shear on the flow is aided by computing the autocorrelation function (ACF) of the velocity and magnetic fields in the $x-y$ plane. This autocorrelation provides a dimensionless of measure of the length or time scale over which the velocity (or magnetic field) maintains a value similar to itself and thus provides a measure of the locality of interactions in a turbulent flow. For random functions, the ACF decays exponentially. A plot of the spatial ACF in the $x-y$ plane characterizes the spatial anisotropy of the velocity and magnetic field fluctuations. 

Following the convention used by \cite{2009ApJ...694.1010G} and \cite{2012MNRAS.422.2685S}, we define the spatial ACF of the magnetic field component `i' ($i=x,y,$ or $z$) as:
\begin{equation}
\text{ACF} (b({\bf \delta x}) ) = \overline{\left (\frac{\sum_i \int b_i ({\bf x + \delta x},t) b_i({\bf x}, t) d^3 {\bf x}} {\int b^2({\bf x}, t) d^3 {\bf x}} \right)},
\label{eq:acfxy}
\end{equation}
where ${b^2=b_x^2 + b_y^2 + b_z^2}$. Note that ACF($b$) is normalized to its maximum value at zero displacement ($\delta x = \delta y = \delta z = 0$). Like \cite{2009ApJ...694.1010G}, we subtract off volume averaged mean quantities ($b = b_{\text{total}} - \langle b \rangle$). The overline represents the time averaging over  $\sim 1000 (1/\Omega)$ time units of the saturated state. We use the analogous definition for the autocorrelation of velocity fields $\text{ACF} (v({\bf \delta x}))$.

Fig. \ref{fig:contour} shows the ACF($v({\bf \delta x})$) and ACF($b({\bf \delta x})$) of the three shear values we study in this paper, $q=1.5, 2.1, 4.2$ for both the hydrodynamic and the magnetohydrodynamic runs. In contrast to previous work on the MRI (e.g., \cite{2009ApJ...694.1010G}, \cite{2012MNRAS.422.2685S}, \cite{2015MNRAS.446.2102N}), the tilt angle observed in plots of ACF($b({\bf \delta x})$) with respect to the y-axis is not constant with respect to variations in $q$. In addition, the hydrodynamic velocity ACF in fig. \ref{fig:contour} for both $q=2.1, 4.2$ is more localized compared to the MHD counterparts at these same $q$. Comparing the MHD ACF plots, the $q=2.1$ and $4.2$ MHD runs show a very localized magnetic field compared to the $q=1.5$ run.

The tilt angles for the  $q<2$ cases previously studied were successfully modeled using an analysis of shear on fluctuations which assumed linear terms dominated nonlinear terms in the Navier-Stokes equation. Given that the $q>2$ cases studied here do not show the same simple monatonic dependencies, we are led to investigate how the ratio of nonlinear to linear terms in the MHD equations vary a function of $q$. In the next section, we will show the non-linear terms in the Navier-Stokes equation do indeed dominate the linear terms for the $q>2$ case when compared to the $q<2$ MRI unstable cases of previous work. This is a step toward identifying  the source of the more subtle dependence of tilt and localization on $q$ in the $q>2$ regime even if though exact dependence cannot yet be predicted analytically.

\subsection{Shear dependence of stress and energy: nonlinearities are more influential for $q>2$ than $q<2$}\label{sec:nonlinear}
Here we provide three lines of evidence consistent with nonlinear terms being more influential than linear terms when it comes to understanding the behavior of stress and energy in saturation as a function of $q$ for the $q>2$ regime compared to the $q<2$ regime. This is why it is more difficult to explain the $q$ trends of tilt angle and localization for the $q>2$ regime than the $q<2$ regime.

\subsubsection{Navier Stokes equation: Explicit comparison of nonlinear vs. linear terms for different $q$ regimes}\label{sec:ns}
\begin{figure}
	\centering
	\includegraphics[width=0.45\textwidth]{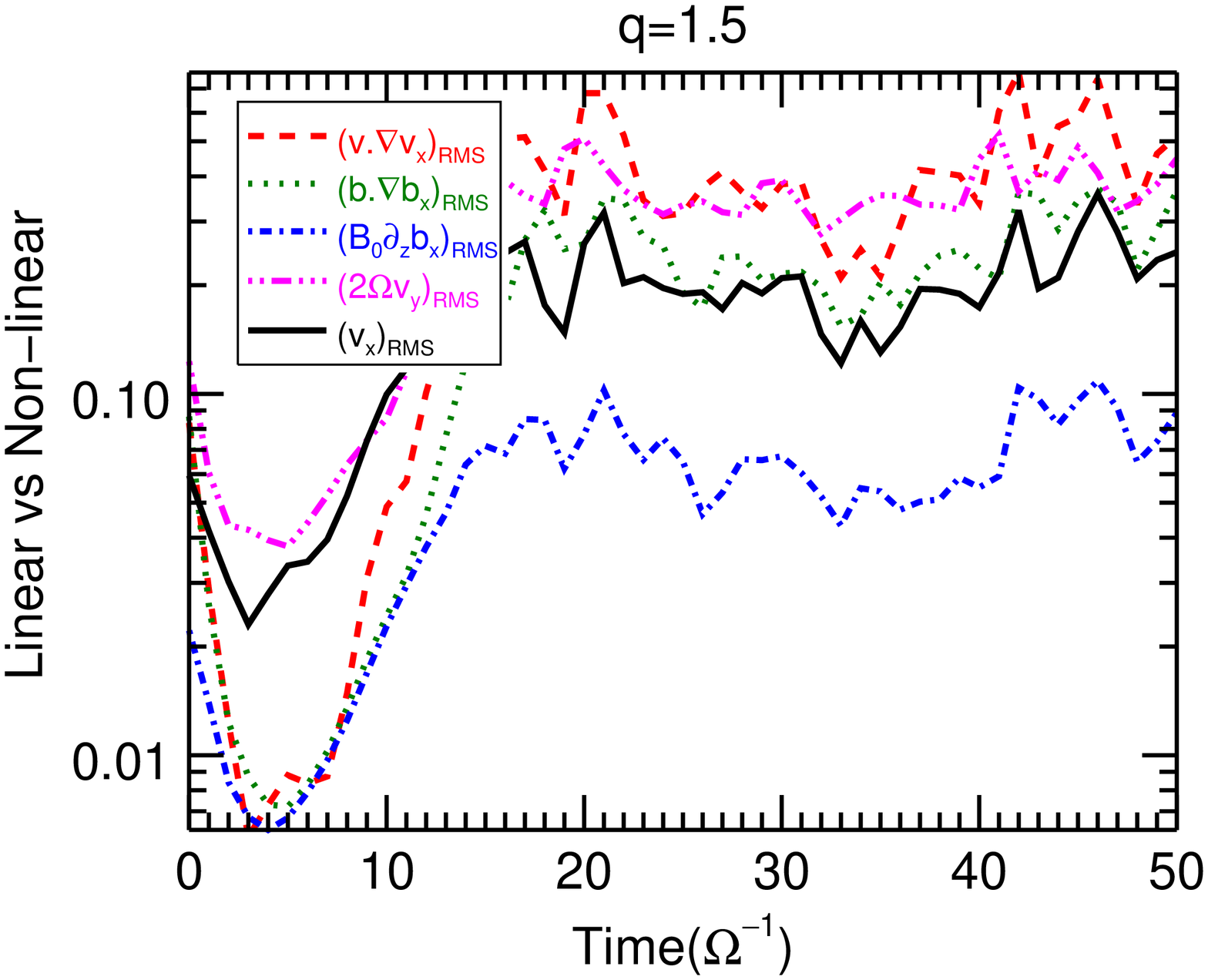}\hspace{10pt}\includegraphics[width=0.45\textwidth]{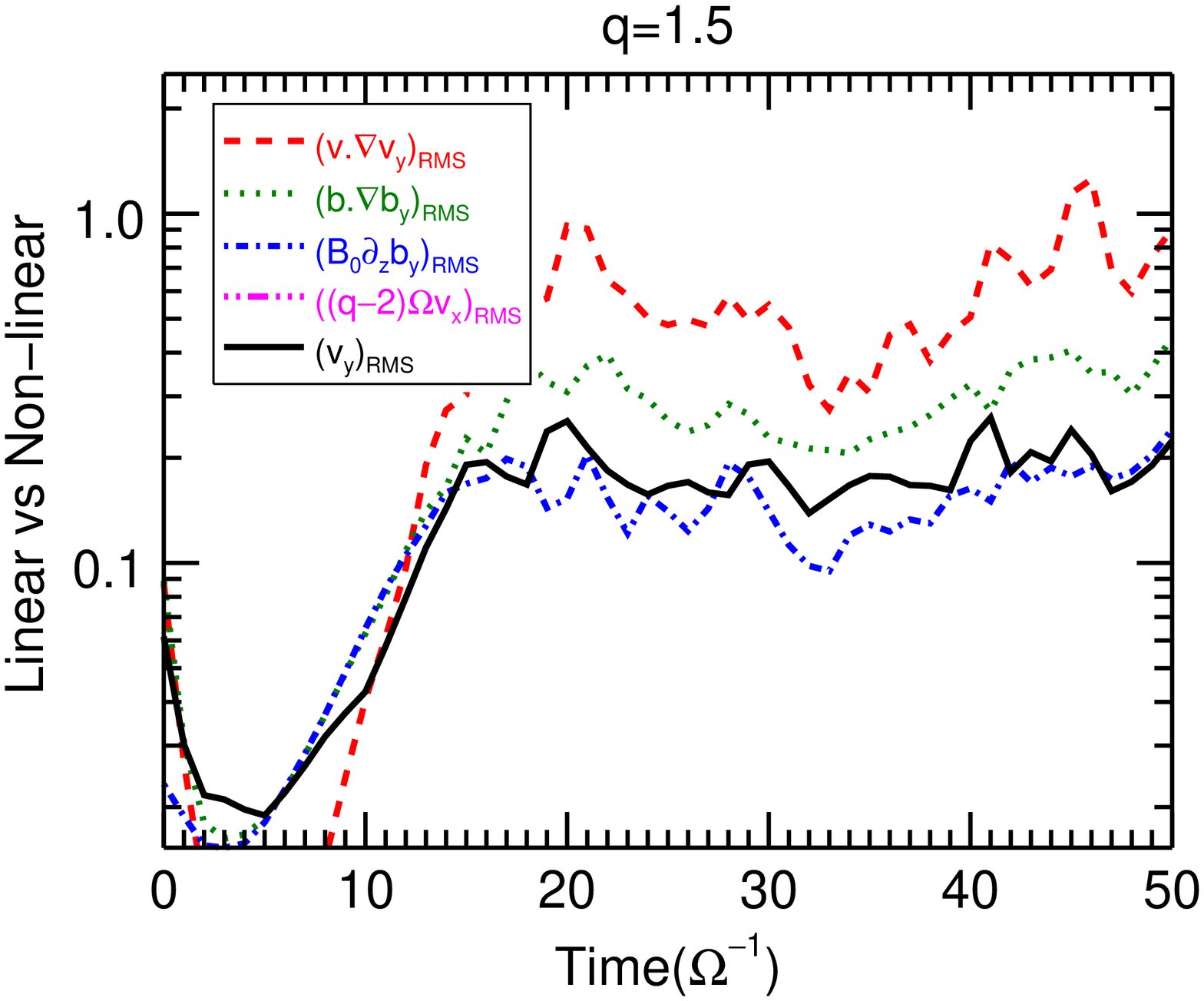} 
	\caption{The comparison of different linear and non-linear terms in eq. \ref{eq:vx} (top panel) and \ref{eq:vy} (bottom panel) for the first 50 $\Omega^{-1}$ times, with $q=1.5$.}
	\label{fig:q15}
\end{figure}

To investigate the effect of shear on the turbulent properties of the flow, we study the time history of the energies and stresses at early times before the flow reaches nonlinear saturation. We focus on the the `x' and `y' velocity equations here:
\begin{align}
\partial_t v_x &= 2\Omega v_y + 
{B_0 \partial_z b_x} + \nu \nabla^2 v_x + \bm{b} \cdot \nabla b_x - \bm{v} \cdot \nabla v_x \label{eq:vx} \\
\partial_t v_y &= (q-2)\Omega v_x +
{B_0 \partial_z b_y} + \nu \nabla^2 v_y + \bm{b} \cdot \nabla b_y - \bm{v} \cdot \nabla v_y \label{eq:vy}.
\end{align}

The last two terms represent non-linear terms in both equations. For $q = 2$, eq. \ref{eq:vy} has no source term in the linear regime and is similar to the (non-rotating) plane Couette flow but with $v_x$ taking the role of shear velocity. In contrast, for $q = 4$ the source terms in eqs. \ref{eq:vx} and \ref{eq:vy} are both proportional to $2\Omega$. The $q = 4$ case results in  apparent isotropy in the two components for the linear regime. 
\begin{figure}
	\centering
	\includegraphics[width=0.45\textwidth]{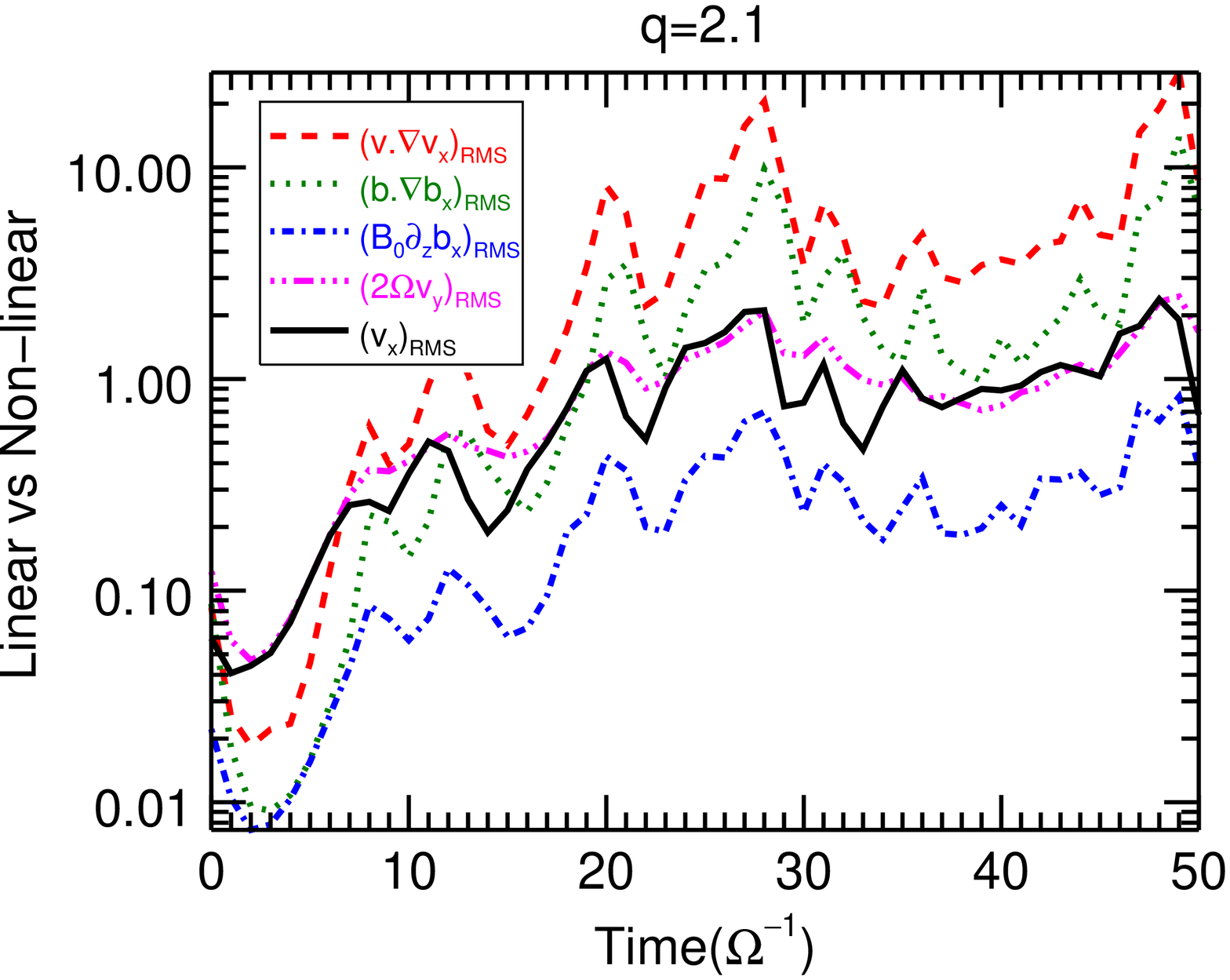}\hspace{10pt}\includegraphics[width=0.45\textwidth]{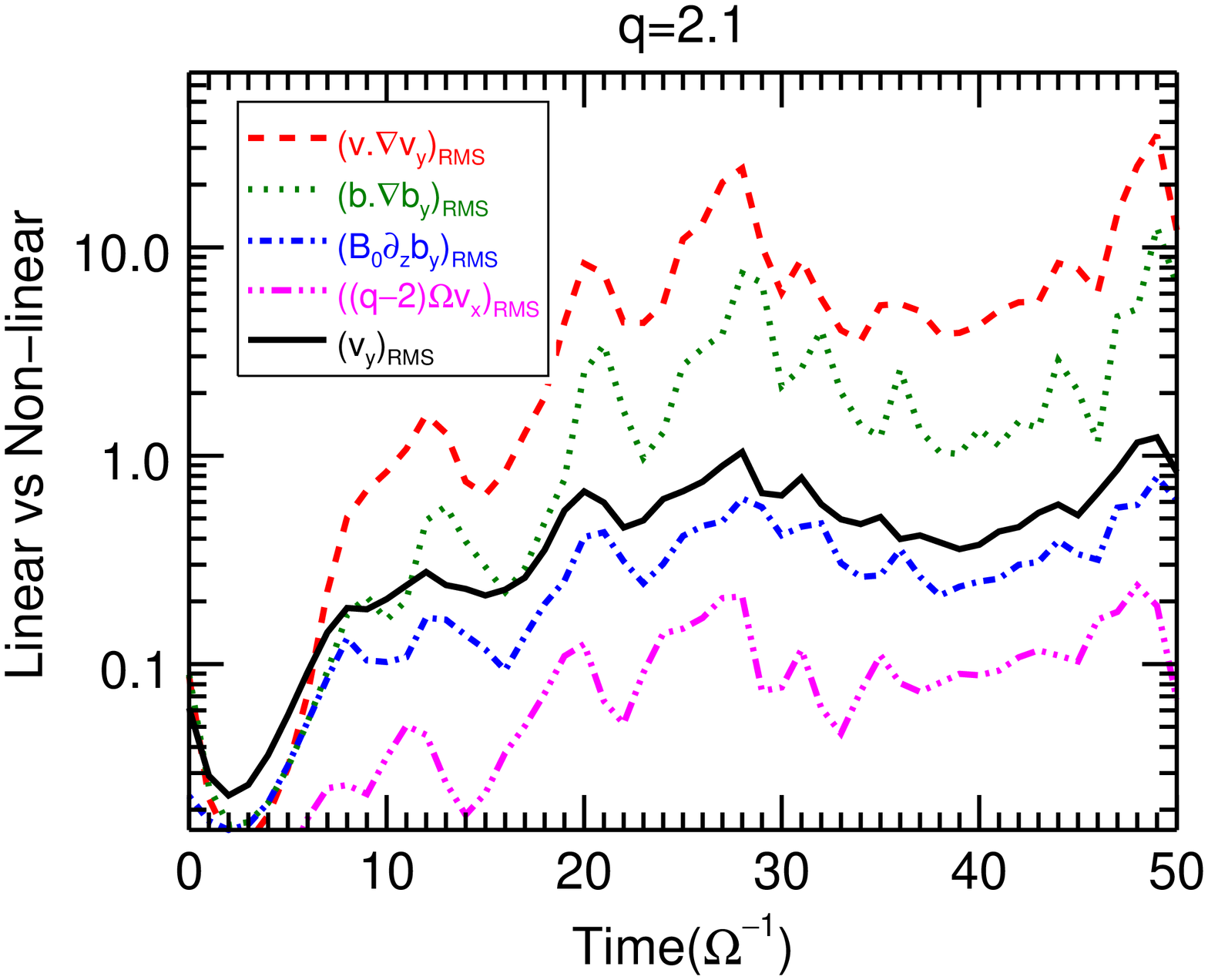} 
	\caption{The comparison of different linear and non-linear terms in eq. \ref{eq:vx} (top panel) and \ref{eq:vy} (bottom panel) for the first 50 $\Omega^{-1}$ times, with $q=2.1$. }
	\label{fig:q21}
\end{figure}
\begin{figure}
	\centering
	\includegraphics[width=0.45\textwidth]{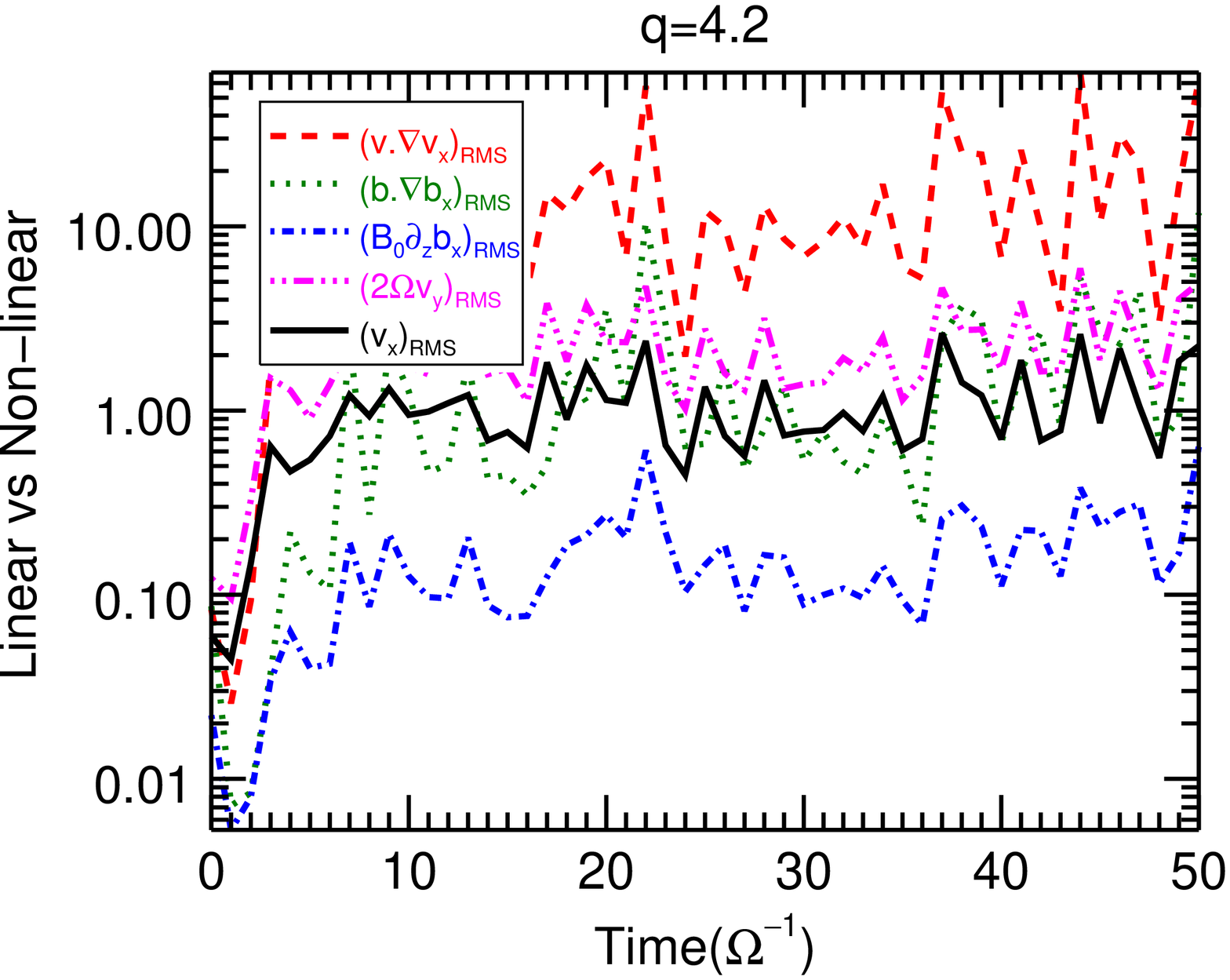}\hspace{10pt}\includegraphics[width=0.45\textwidth]{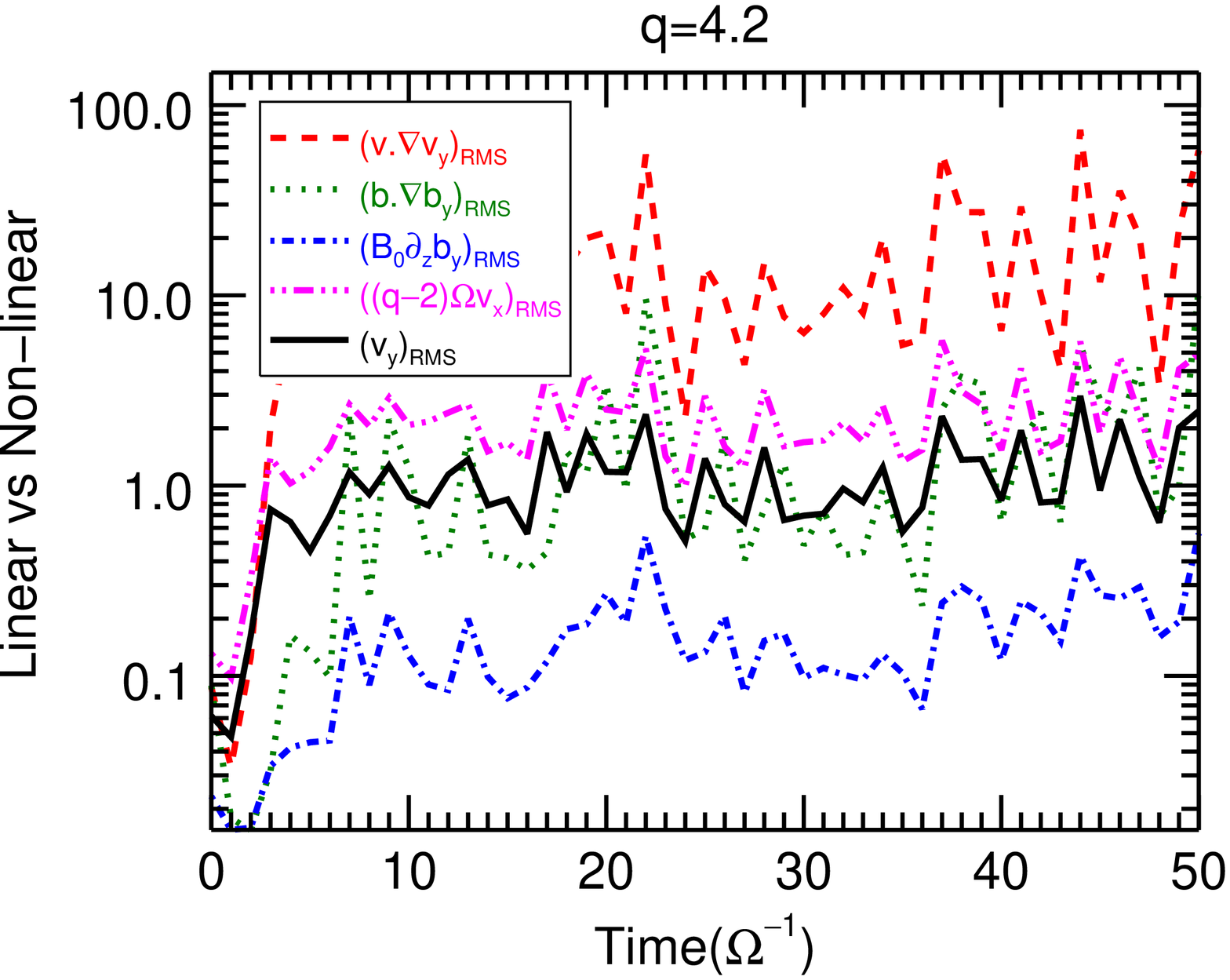} 
	\caption{The comparison of different linear and non-linear terms in eq. \ref{eq:vx} (top panel) and \ref{eq:vy} (bottom panel) for the first 50 $\Omega^{-1}$ times, with $q=4.2$.}
	\label{fig:q42}
\end{figure}

We plot the evolution of the different linear terms in the two equations and compare them with the rms value of the non-linear terms $\bm{v} \cdot \nabla \bm{v}$ and $\bm{b} \cdot \nabla \bm{b}$ for early times  first $20 \Omega^{-1}$ times in figs. \ref{fig:q15}, \ref{fig:q21}, \ref{fig:q42}.  

For $q=1.5$, the  $2\Omega v_y$ term in eq. \ref{eq:vx} is comparable to the non-linear terms for $q=1.5$ (top left panel of fig. \ref{fig:q15}), suggesting that for $q<2$ the linear effects are very influential even as the saturated state is approached. This is assessed visually by noting that the red dashed curve overshoots the magnetic curve at most in the last few time steps of this plot. The linear term due to magnetic tension, $B_0 \partial_z b_x$\footnote{For an initially zero net flux case, such a term would be absent in the linear limit. We did carry out zero net flux simulations for $B_{\text{z,ini}} = B_0 \sin k_x x$ for all three shear values at $Re=Rm=1600$ and found that only the $q=4.2$ run shows growth and sustenance of kinetic and magnetic energy while for the other two runs, both kinetic and magnetic energy decay.},  is nearly an order of magnitude weaker than the other terms in this plot.

For $q>2$ the top panels of (figs. \ref{fig:q21}, \ref{fig:q42}), show that the corresponding  linear terms are nearly an order of magnitude weaker than nonlinear terms \ref{eq:vx}. Note here that the red dashed curve dominates over a longer range of time compared to the $q=1.5$ case. Since the non-linear effects are dominating the linear velocity and magnetic field terms in this regime, the flow in this regime is expected to be more random with a smaller correlation length, consistent with fig. \ref{fig:contour}. Note also that for $q>2$ (particularly in the $q=4.2$ plot) the non-linear magnetic terms $\bm{b} \cdot \nabla b_i$ (where $i=x$ or $y$) are considerably weaker than the corresponding non-linear velocity term $\bm{v} \cdot \nabla v_i$ (red dashed), suggesting that magnetic effects are subdominant in both the linear and non-linear regimes for the $q>2$ regime (eq. \ref{eq:vx}).

Analogously, comparing the linear vs nonlinear terms of \ref{eq:vy} for $q=1.5$ vs $q>2$ we find that in this case the nonlinear terms dominate the linear terms in both regimes, but that the red dashed curves of the bottom panels of figs. \ref{fig:q21} and \ref{fig:q42} are more dominant over a longer time range than in the bottom panel of fig. \ref{fig:q15}.

\subsubsection{Induction equation: Explicit comparison of nonlinear vs. linear terms for different $q$ regimes}\label{sec:ind}
The induction equation has the form:
\begin{align}
\partial_t b_x &= {B_0 \partial_z v_x} + \eta \nabla^2 b_x + \bm{b} \cdot \nabla v_x - \bm{v} \cdot \nabla b_x \label{eq:bx} \\
\partial_t b_y &= -q\Omega b_x +
{B_0 \partial_z b_y} + \eta \nabla^2 b_y + \bm{b} \cdot \nabla v_y - \bm{v} \cdot \nabla b_y \label{eq:by}.
\end{align}

\begin{figure}
	\centering
	\includegraphics[width=0.45\textwidth]{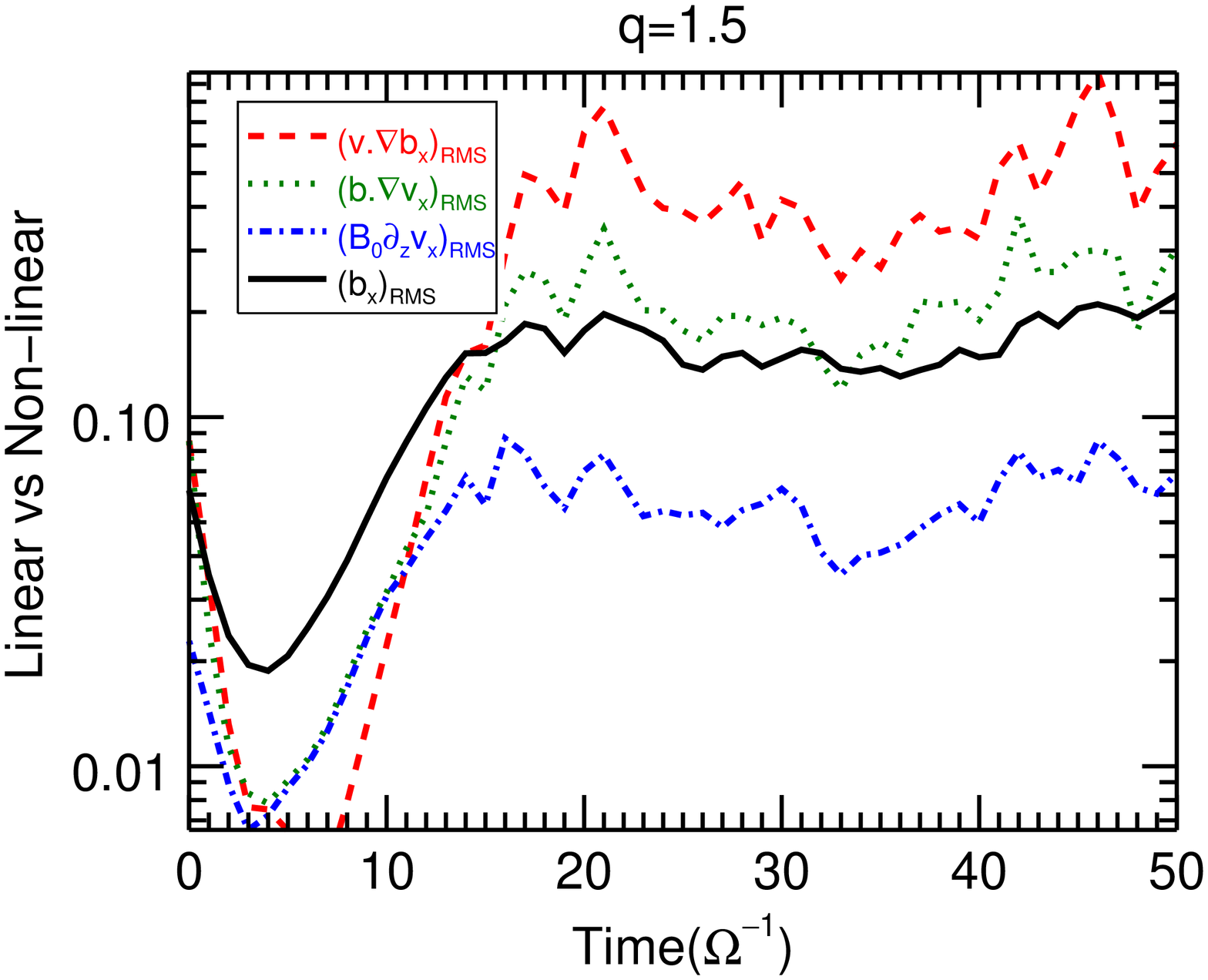}\hspace{10pt}\includegraphics[width=0.45\textwidth]{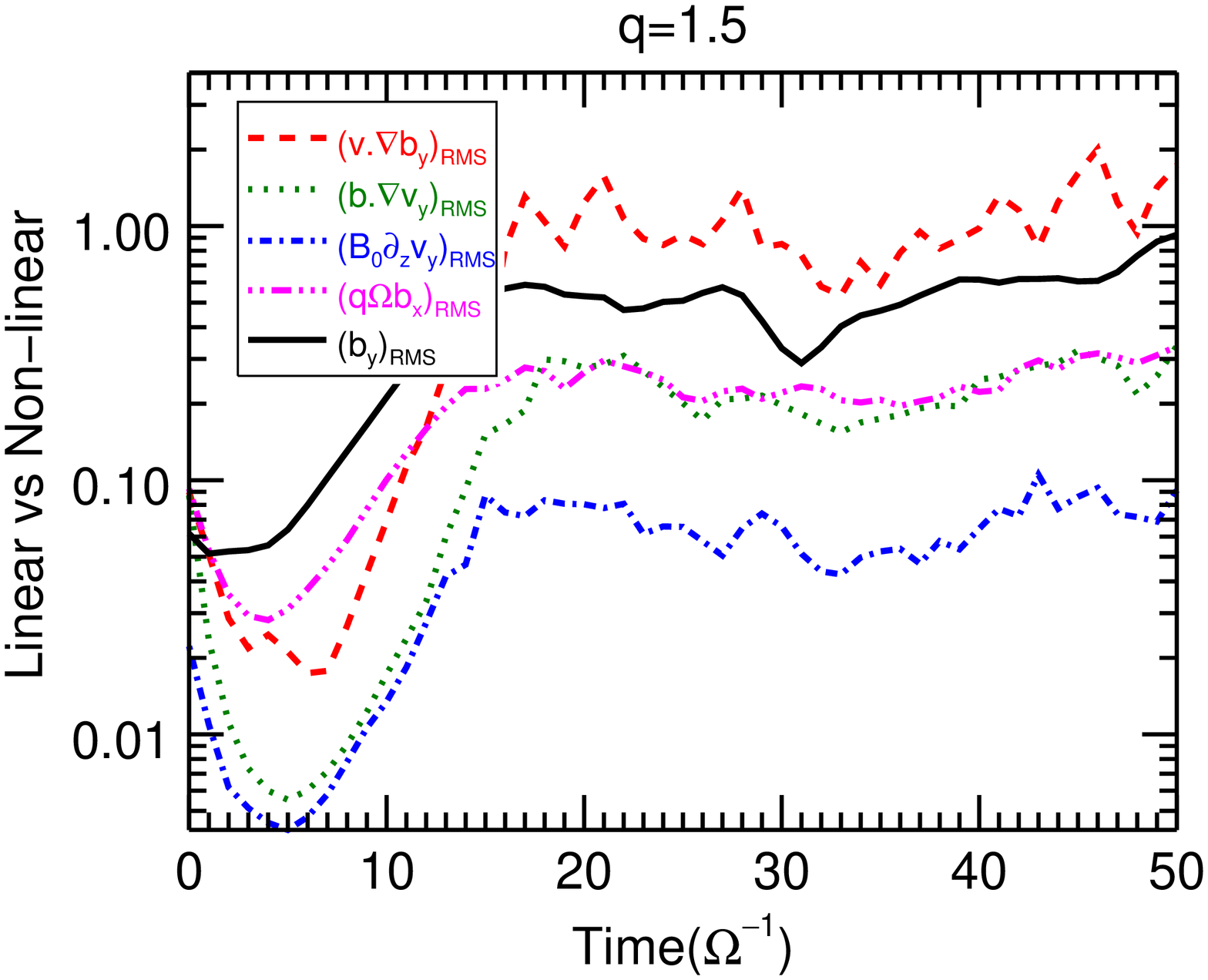} 
	\caption{The comparison of different linear and non-linear terms in eq. \ref{eq:bx} (top panel) and \ref{eq:by} (bottom panel) for the first 50 $\Omega^{-1}$ times, with $q=1.5$.}
	\label{fig:q15b}
\end{figure}
\begin{figure}
	\centering
	\includegraphics[width=0.45\textwidth]{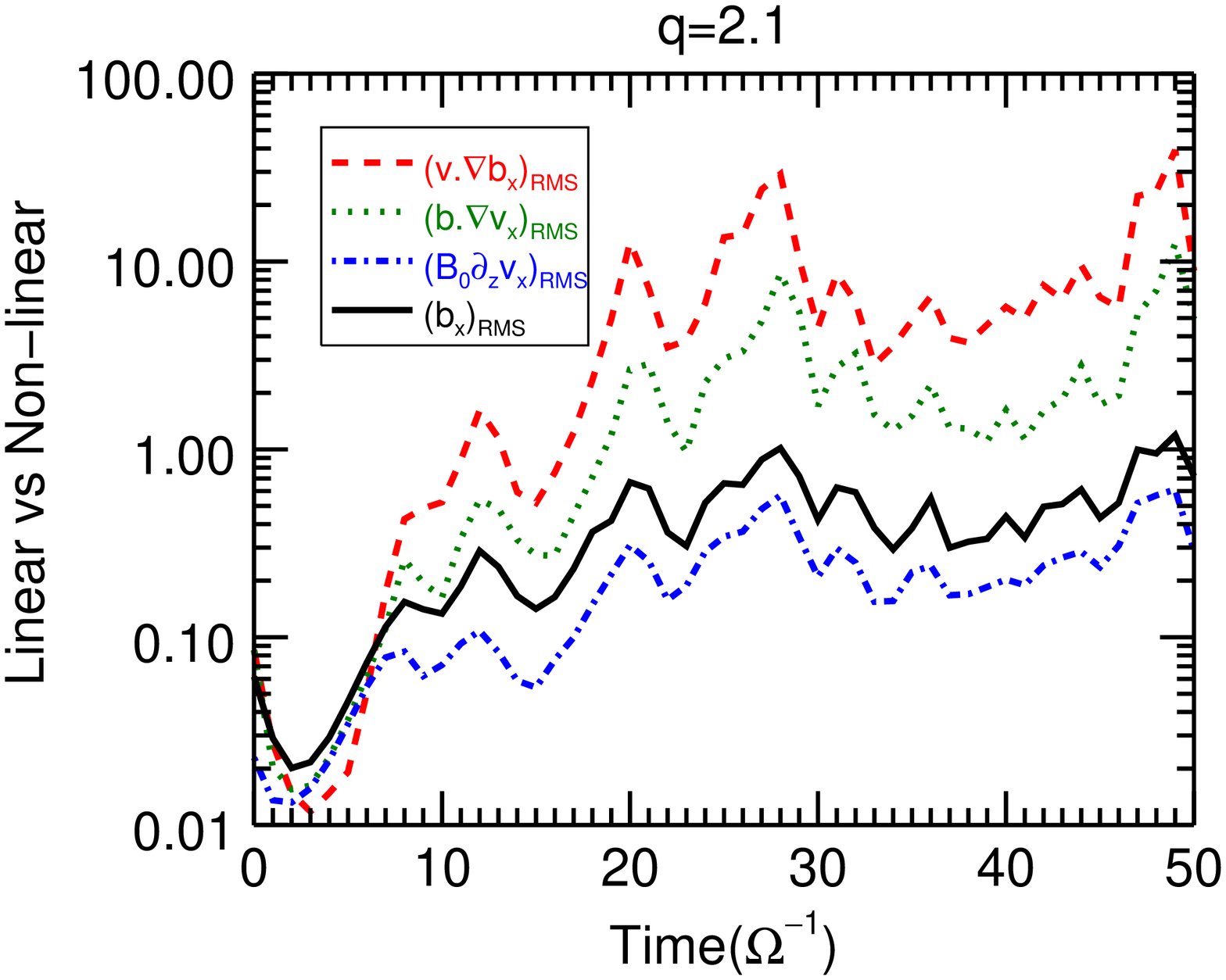}\hspace{10pt}\includegraphics[width=0.45\textwidth]{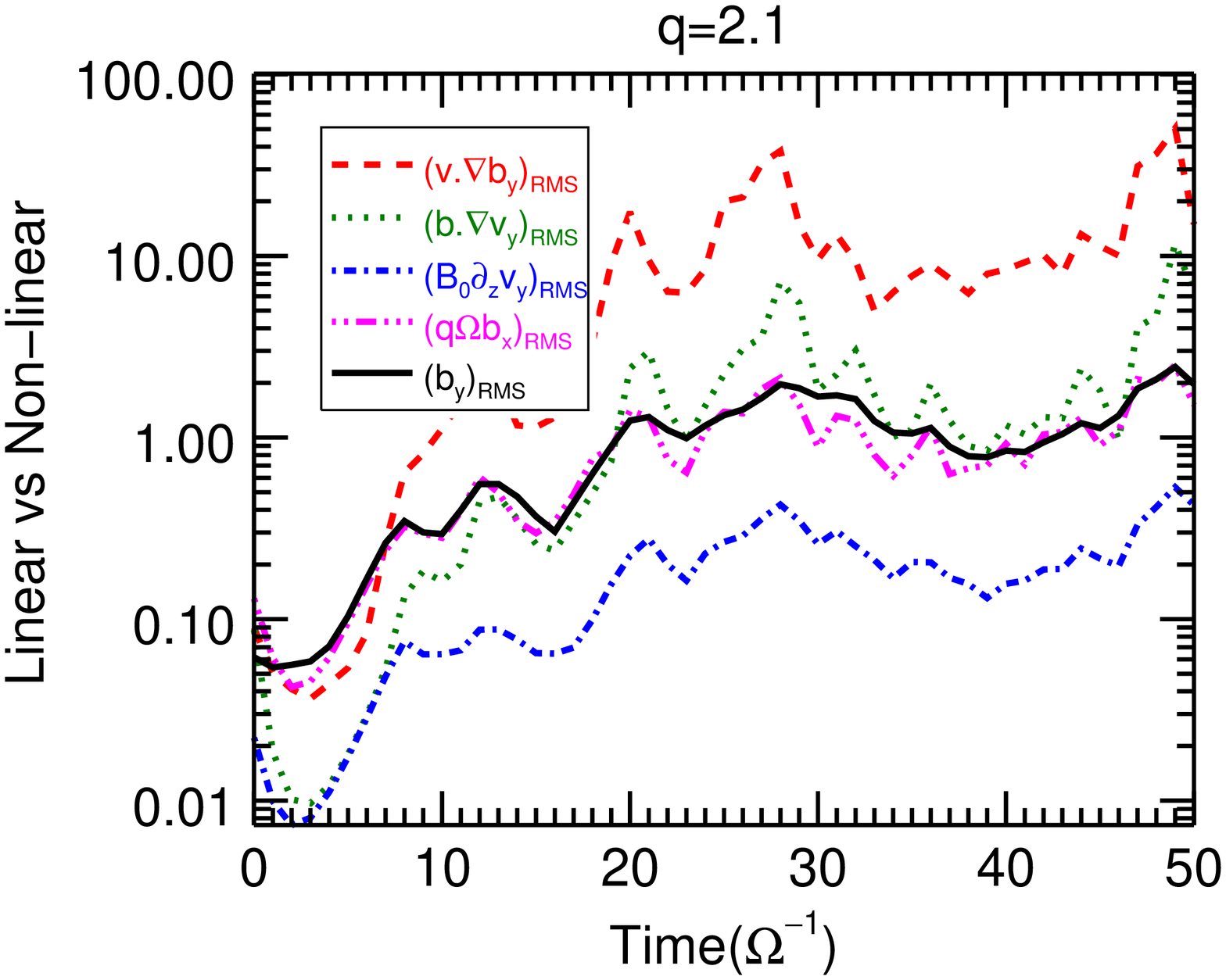} 
	\caption{The comparison of different linear and non-linear terms in eq. \ref{eq:bx} (top panel) and \ref{eq:by} (bottom panel) for the first 50 $\Omega^{-1}$ times, with $q=2.1$.}
	\label{fig:q21b}
\end{figure}
\begin{figure}
	\centering
	\includegraphics[width=0.45\textwidth]{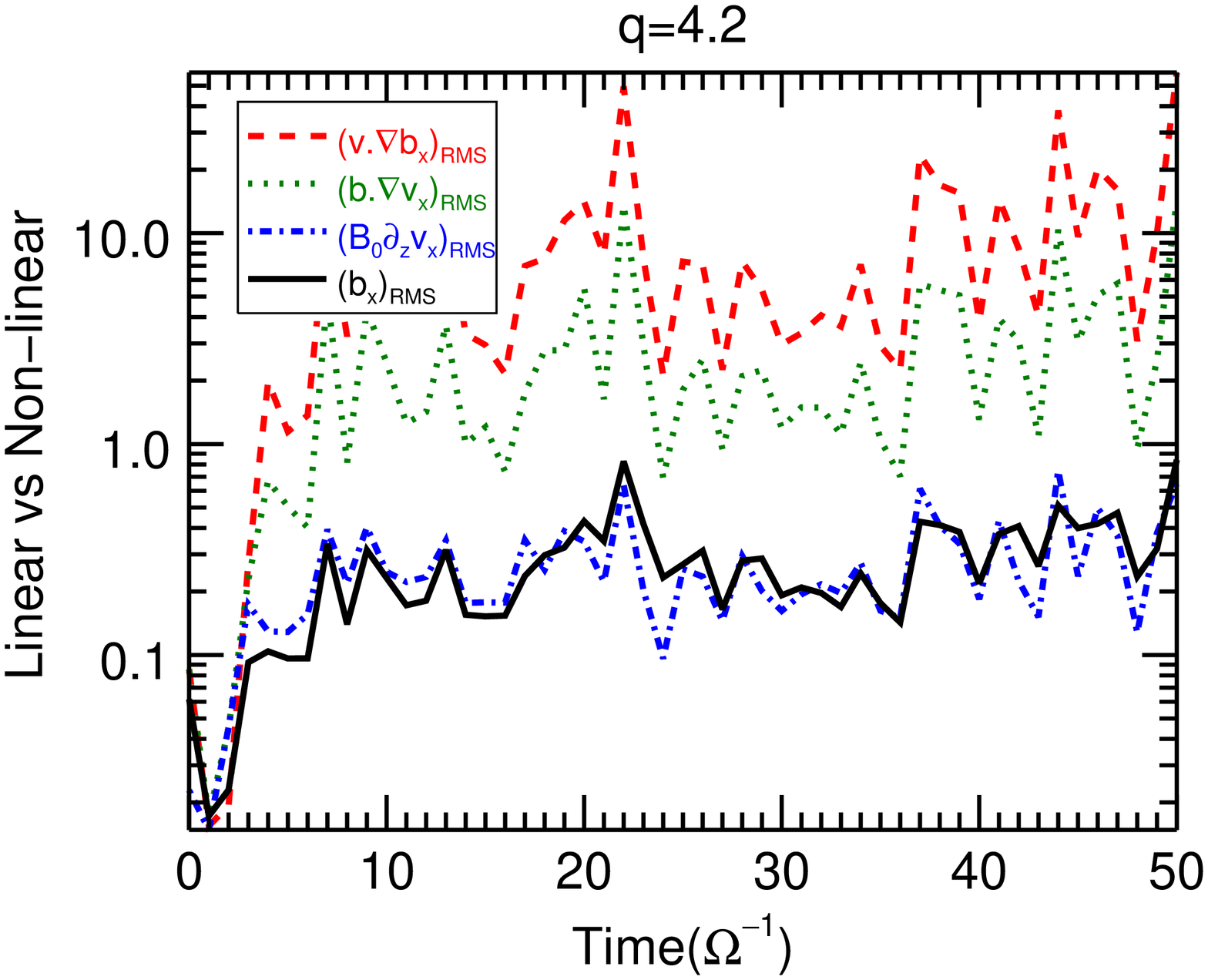}\hspace{10pt}\includegraphics[width=0.45\textwidth]{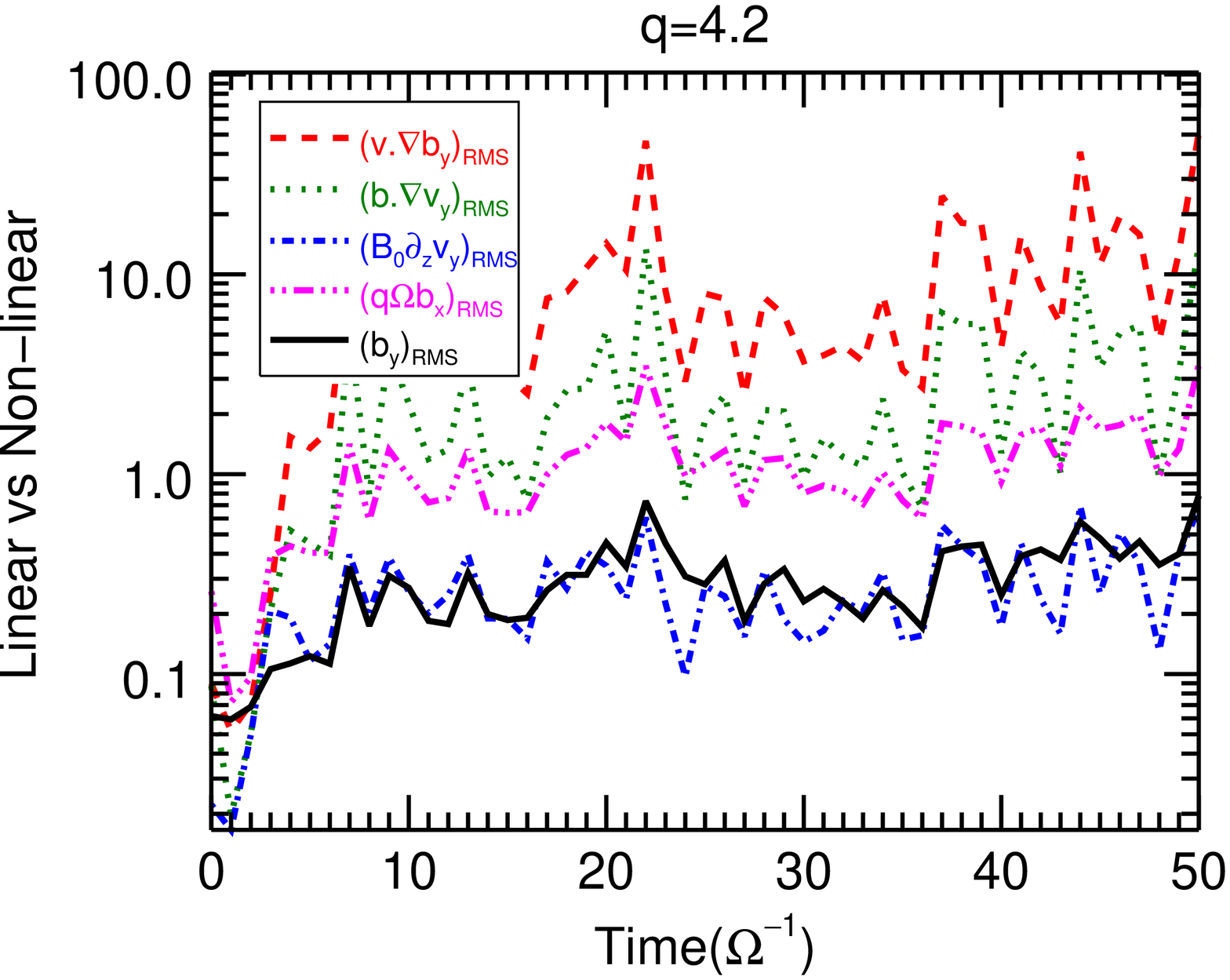} 
	\caption{The comparison of different linear and non-linear terms in eq. \ref{eq:bx} (top panel) and \ref{eq:by} (bottom panel) for the first 50 $\Omega^{-1}$ times, with $q=4.2$.}
	\label{fig:q42b}
\end{figure}

The first two terms in the $b_x$ equation (eq. \ref{eq:bx}) and the first three terms in the $b_y$ equation (eq. \ref{eq:by}) are linear. The terms of the form $\bm{v}\cdot\nabla\bm{b}$ and $\bm{b}\cdot\nabla\bm{v}$ are nonlinear because the velocity fields depend on the magnetic fields through the Navier Stokes equation (eqs. \ref{eq:vx} and \ref{eq:vy}). When the magnetic fields are weak $b^2 \ll v^2$, then these terms could be considered approximately linear. However, for all of the shear values considered in this paper, the magnetic and kinetic energy are comparable right from the beginning of the simulations so it appears that the last two terms in both eqs. \ref{eq:bx} and \ref{eq:by} are nonlinear. 

The bottom panel in figures \ref{fig:q15b}, \ref{fig:q21b} and \ref{fig:q42b} show that the generation of the azimuthal field $b_y$ due to the shearing of the radial field $b_x$ is very significant in the first few rotation times $\Omega^{-1}$ but is nearly an order of magnitude weaker than the $\bm{v}\cdot\nabla b_y$ term in the saturated regime. The other nonlinear term $\bm{b}\cdot\nabla v_y$ is slightly larger in magnitude for $q=2.1$ and $q=4.2$ compared to the $q\Omega b_x$ terms in the saturation regime but the two terms are nearly equal for $q=1.5$. This suggests that stretching is more important for field growth in the $q>2$ regime than the $q<2$ regime.

\subsubsection{Dependence of stresses and correlation time on $q$}\label{sec:corrtime}

\begin{figure}
\centering
		\includegraphics[width=0.45\textwidth]{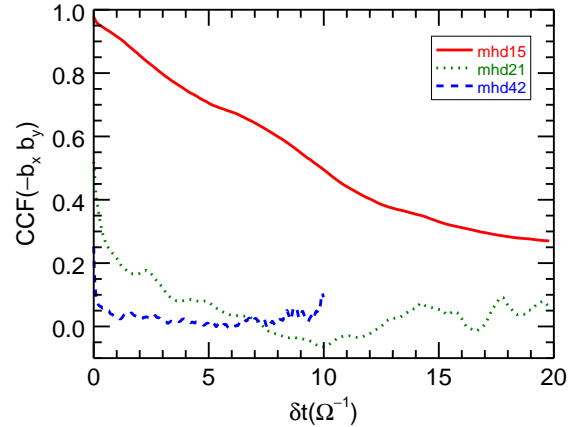}
		\caption{The CCF($b_xb_y(\delta t)$) as defined in eq. \ref{eq:ccfT} but only for MHD runs. The x-axis is in units of $1/\Omega$. The colour scheme is as follows: mhd15 (red), mhd21 (green), mhd42 (blue).}
		\label{fig:bxbyvsq}
\end{figure}
\begin{figure}
	\centering
		\includegraphics[width=0.45\textwidth]{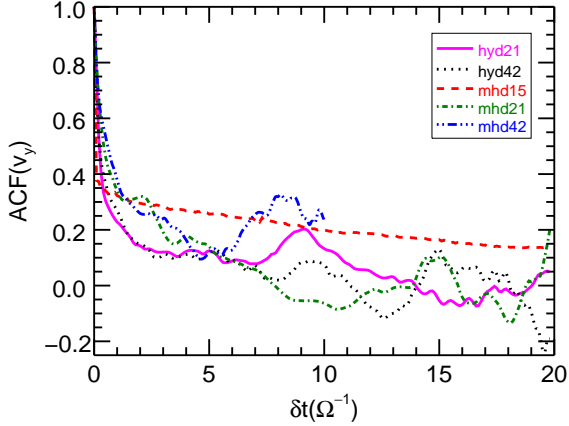}
		\caption{The ACF($v_y(\delta t)$) as defined in eq. \ref{eq:acfT}. The colour scheme is as follows: mhd15 (red), mhd21 (green), mhd42 (blue), hyd21 (magenta), hyd42 (black).}
		\label{fig:vyvsq}
\end{figure}
\begin{figure}
		\includegraphics[width=0.45\textwidth]{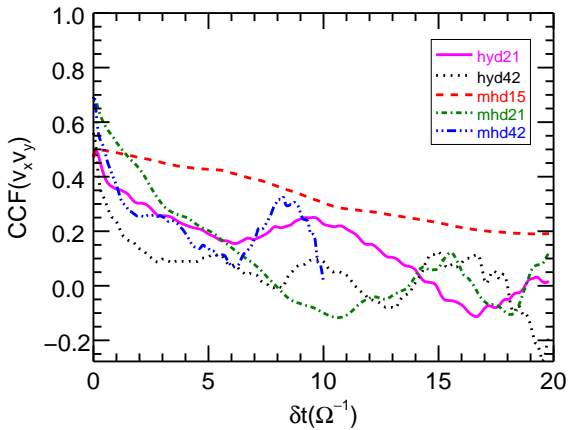}
		\caption{The CCF($v_xv_y(\delta t)$) as defined in eq. \ref{eq:ccfT}. Colour scheme same as fig. \ref{fig:vyvsq}.}
		\label{fig:vxvyvsq}
\end{figure}

To evaluate how $\alpha_{\text{kin,y}} (\equiv \overline{\langle v_xv_y\rangle/\langle v_y^2\rangle})$ and $\alpha_{\text{mag,y}} (= \overline{-\langle b_xb_y \rangle/\langle b_y^2 \rangle})$ vary with shear, we use the autocorrelation function of time to obtain the correlation time. Following our earlier work \citep{2015MNRAS.446.2102N}:
\begin{equation}
\text{ACF} (v_y(\delta t)) = \overline{\left( \frac{\int v_y (\bm{x},t + \delta t) v_y( \bm{x}, t) dt} {\int v_y^2(\bm {x}, t) dt} \right)}
\label{eq:acfT}
\end{equation}
where the angle brackets represent volume averaging over all space. Time integration is done over several orbits in the turbulent saturated state. Similarly we can calculate the cross correlation in time of $b_x$ with $b_y$ and $v_x$ and $v_y$. For example, for the velocities we have:
\begin{equation}
\text{CCF} (v_xv_y(\delta t)) = \overline{\left( \frac{\int v_x (\bm{x},t + \delta t) v_y( \bm{x}, t) dt} {\sqrt{\int v_x^2(\bm {x}, t) dt \int v_y^2(\bm {x}, t) dt}}\right)}.
\label{eq:ccfT}
\end{equation}

The correlation times, computed from an exponential fits to the plot of the ACF or CCF as in Fig. \ref{fig:bxbyvsq}, tells us the characteristic time scale over which turbulent quantities such as the velocity are correlated to themselves or other quantities. From MRI simulations with $q<2$, \cite{2015MNRAS.446.2102N} found that the correlation time between $x$ and $y$ components of the field $\tau$ was roughly inversely proportional to the shear. There  the stress ACF was calculated instead of the CCF but we checked that the CCF exhibits a similar $1/q$ behavior in the $q<2$ regime \cite{shearerr}. 

The importance of the correlation time is that when linear stretching in the induction equation can be used to estimate the amplification of azimuthal fluctuations from radial fluctuations, the azimuthal field is amplified by shear during a correlation time with dominant term
\begin{equation}
\alpha_{\text{mag,y}} = -\langle b_x b_y\rangle/ \langle b_y^2\rangle \sim |q\Omega| \tau
\label{eq:bystress}
\end{equation}
If $\tau \sim 1/q\Omega$, $\alpha_{\text{mag,y}}$ is roughly constant with shear. Indeed for $q<2$ case, this was confirmed by the simulations.  Moreover, the correlation times for the three quantities $v_y$, $b_x$ and $b_xb_y$ were very similar (see figure 13 of \cite{2015MNRAS.446.2102N}). Using a similar argument for velocity as in eq. ~\ref{eq:bystress}, we would get
\begin{equation}
\alpha_{\text{kin,y}} = \lb v_xv_y \rb / \lb v_y^2 \rb \sim (|(q-2)\Omega|\tau)^{-1}.
  \label{eq:vyy}
\end{equation}

We now assess whether the above two equations, which are rooted in linear analysis, are equally effective at explaining the trends found in the $q>2$ cases. We focus on the CCF (which is more relevant that the ACF) for stresses.
\begin{figure}
	\centering
	\includegraphics[width=0.45\textwidth]{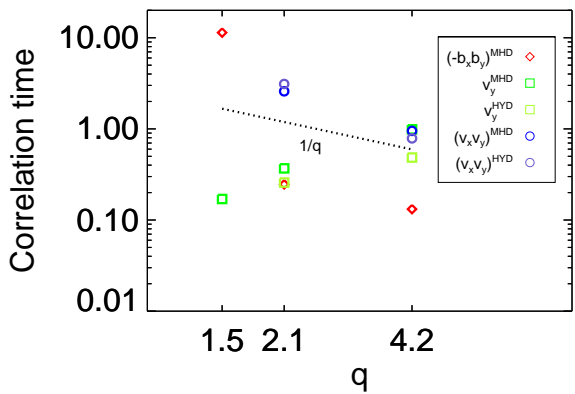}
	\caption{Correlation time calculated from an exponential fit to the MHD simulation plots in figs. \ref{fig:vyvsq}, \ref{fig:vxvyvsq}, \ref{fig:bxbyvsq}. The y-axis is in units of $1/\Omega$. }
	\label{fig:vystresses}
\end{figure}
We find that $\alpha_{\text{mag,y}}$ is nearly constant just like the $q<2$ MRI regime (see table \ref{tab:1}) owing to the $1/q$ dependence of the correlation time for CCF ($b_xb_y(\delta t)$) (fig. \ref{fig:bxbyvsq}). However, $\alpha_{\text{kin,y}}$ decreases both for the HD and MHD runs unlike the $q<2$ cases\footnote{Fig. 7 of \cite{2015MNRAS.446.2102N} shows that $\lb v_xv_y\rb/\lb v^2\rb$ increases with shear. We did not plot the CCF ($v_xv_y(\delta t)$) in that paper but we checked that the correlation time for the Reynolds stress also varies as $1/q$, which in a linear picture would explain the increase in the ratio of Reynolds stress to the kinetic energy as eq. \ref{eq:vyy} suggests with the assumption $\lb v^2\rb \sim \lb v_y^2\rb$.} Eq. (\ref{eq:vyy}) would require that for $\alpha_{\text{kin,y}}$ to decrease, $\tau$ has to go down faster than $|q-2|^{-1}$. To check this, we plot the ACF of $v_y$ in fig. ~\ref{fig:vyvsq}, which shows a slight increase with shear. Then $\tau$ would be predicted to decrease by a factor of more than 22 as $q$ varies from $q=2.1$ to $q=4.2$ if Eq. (\ref{eq:vyy}) were the whole story. But the CCF of $v_xv_y$ in fig. ~\ref{fig:vxvyvsq} shows only a factor of $3$ (for MHD) to $4$ (for HD) times decrease with shear (see fig. ~\ref{fig:vystresses}). The likely explanation for this  discrepancy is that Eq. (\ref{eq:vyy}) does not capture the effect of nonlinear terms. Indeed the comparison of linear and non-linear terms in figs. \ref{fig:q21}, \ref{fig:q42} shows that non-linear terms are generally more important than the linear terms for $q>2$ regime.

\subsubsection{Tilt angle dependence on $q$}\label{sec:tilt}
The tilt angle in ACF($b({\bf \delta x})$) (fig. \ref{fig:contour}) has been directly connected to the ratio of  Maxwell stress to magnetic energy $\langle -b_xb_y \rangle/\langle b^2 \rangle$ in previous work on MRI (e.g. \cite{2015MNRAS.446.2102N}). Here we modify the definition to compare the stress to just the y-component of magnetic field squared, $\alpha_{\text{mag},y} = -\langle b_xb_y \rangle/\langle b_y^2 \rangle = \tan \theta_{\text{tilt}}$. 

For our Rayleigh unstable simulations, the tilt angle observed from the ACF($b({\bf \delta x})$) and the definition based on $\alpha_{\text{mag},y}$ \footnote{We thank the referee for pointing this out.} disagree, in contrast the MRI $q<2$ cases. For example, for $q=2.1$ the $\alpha_{\text{mag},y} = 0.2334$ which is equivalent to $\theta_{\text{tilt}} \sim 13.14^{\circ}$ whereas for $q=4.2$, the $\alpha_{\text{mag},y} = 0.1712$ which is equivalent to $\theta_{\text{tilt}} \sim 9.71^{\circ}$ (fig. ~ \ref{fig:contour}). A visual inspection of fig. ~\ref{fig:contour} shows that the $q=4.2$ tilt angle is nearly $45^{\circ}$. 

From our discussion in sections \ref{sec:ns} and \ref{sec:ind}, we are led again to the conclusions that this is further evidence for the more dominant role of nonlinear terms in the Rayleigh unstable regime compared to the MRI unstable $q<2$ regime. This demonstrates the inadequacy of linear arguments to explain the correlation between $b_x$ and $b_y$.
 
At present we do not have a non-linear model to explain the observed behavior in either the spatial correlation (fig. \ref{fig:contour}) and the temporal correlation (fig. \ref{fig:vystresses}) but the identification that the nonlinear terms are essential is a step toward such. The importance of these nonlinear terms present a challenge for analytic explanations.

\section{Conclusions}\label{sec:conclusions}
We have compared the turbulent saturation properties of Rayleigh unstable MHD shear flows with those of the more commonly studied MRI unstable but Rayleigh stable regime. Our results are summarized below:
\begin{enumerate}
	\item The Rayleigh unstable regime ($q>2$) generates turbulent velocity flows with or without magnetic fields. In the presence of magnetic fields, the fluid turbulence drives dynamo amplification of the total magnetic energy.
	\item In this $q>2$ regime, we find that magnetic energy and Maxwell stresses saturate at lower values than the kinetic energy fluctuations and associated Reynolds stresses. In this regime therefore, the magnetic field is ``slaved" to the flow turbulence. This contrasts the MRI unstable regime ($q<2$) in which the magnetic fluctuations and magnetic stresses dominate the kinetic energy fluctuations and stresses.
    \item The quantity $\alpha_{\text{mag,y}}$ remains roughly constant in $q>2$ regime, which is the same as for the $q<2$ regime (\cite{2015MNRAS.446.2102N}). The tilt angle in ACF($b({\bf \delta x})$), on the other hand, with respect to y-axis is not constant as $q$ changes. This contrasts the behavior in the MRI regime where the tilt angle is constant with changing $q$.
  	\item We found that the magnetic structures of the flow become more localized as we increase the shear from $q=1.5$ to $4.2$. 
\end{enumerate}

Our work on MHD turbulence in the Rayleigh unstable regime has shown qualitative differences in the way quantities scale with $q$ compared to  the more well studied MRI unstable regimes. While the dependencies on $q$ for MRI regime seems to be captured by analytic explanations that invoke linear analysis, the same linear estimates do not work for the $q>2$ cases. We have traced the source of these differences to the stronger influence of non-linear effects in the Rayleigh unstable regime. A physical and analytic understanding of these differences requires non-linear modeling of MHD shear turbulence in the two regimes, which is good opportunity for work beyond the present scope.

\bibliography{general}
\bibliographystyle{mnras}

\section*{Acknowledgments}
We thank G. Lesur for discussions about the \textsc{snoopy} code. FN acknowledges Horton Fellowship from the Laboratory for Laser Energetics at U. Rochester. We acknowledge support from NSF grant AST-1109285. EB acknowledges support from the Simons Foundation and the IBM-Einstein Fellowship fund while at IAS, and grants HST-AR-13916.002 and NSF-AST1515648.  We acknowledge the Center for Integrated Research Computing at the University of Rochester for providing computational resources.

\bsp

\label{lastpage}

\end{document}